# GPU-based Monte Carlo radiotherapy dose calculation using phase-space sources

**Reid W Townson[1,2,3], Xun Jia[3], Zhen Tian[3], Yan Jiang Graves[3], Sergei Zavgorodni[2,1] and Steve B Jiang[3]**

[1] Department of Physics and Astronomy, University of Victoria, PO Box 3055, STN CSC, Victoria, British Columbia V8W 3P6, Canada

[2] Department of Medical Physics, BC Cancer Agency, Vancouver Island Centre, 2410 Lee Avenue, Victoria, British Columbia V8R 6V5, Canada

[3] Center for Advanced Radiotherapy Technologies and Department of Radiation Medicine and Applied Sciences, University of California San Diego, La Jolla, CA 92093, US

Email: rtownson@uvic.ca, sbjiang@ucsd.edu

**Abstract**

A novel phase-space source implementation has been designed for GPU-based Monte Carlo dose calculation engines. Short of full simulation of the linac head, using a phase-space source is the most accurate method to model a clinical radiation beam in dose calculations. However, in GPU-based Monte Carlo dose calculations where the computation efficiency is very high, the time required to read and process a large phase-space file becomes comparable to the particle transport time. Moreover, due to the parallelized nature of GPU hardware, it is essential to simultaneously transport particles of the same type and similar energies but separated spatially to yield a high efficiency. We present three methods for phase-space implementation that have been integrated into the most recent version of the GPU-based Monte Carlo radiotherapy dose calculation package gDPM v3.0. The first method is to sequentially read particles from a patient-dependent phase-space and sort them on-the-fly based on particle type and energy. The second method supplements this with a simple secondary collimator model and fluence map implementation so that patient-independent phase-space sources can be used. Finally, as the third method (called the phase-space-let, or PSL, method) we introduce a novel source implementation utilizing pre-processed patient-independent phase-spaces that are sorted by particle type, energy and position. Position bins located outside a rectangular region of interest enclosing the treatment field are ignored, substantially decreasing simulation time with little effect on the final dose distribution. The three methods were validated in absolute dose against BEAMnrc/DOSXYZnrc and compared using gamma-index tests (2%/2mm above the 10% isodose). It was found that the PSL method has the optimal balance between accuracy and efficiency and thus is used as the default method in gDPM v3.0. Using the PSL method, open fields of 4x4, 10x10 and 30x30 $cm^2$ in water resulted in gamma passing rates of 99.96%, 99.92% and 98.66%, respectively. Relative output factors agreed within 1%. An IMRT patient plan using the PSL method resulted in a passing rate of 97%, and was calculated in 50 seconds (per GPU) compared to 8.4 hours (per CPU) for BEAMnrc/DOSXYZnrc.

## 1. Introduction

Monte Carlo (MC) simulation is well recognized as the most accurate method for radiotherapy dose calculation.

Clinical implementation of MC algorithms in commercial systems to date has been limited due to long computation times and substantial hardware requirements. In recent years, dose calculation engines that utilize the graphics processing unit (GPU) have been developed by several groups and demonstrated to have significant speed benefits over classical (CPU-based) codes (Badal and Badano 2009, Hissoiny *et al.* 2011, Jia *et al.* 2011, Pratx and Xing 2011, Jahnke *et al.* 2012).

The source model of a linear accelerator (linac) head in a MC dose calculation plays a major role in the accuracy of the resulting dose distributions. A good characterization of a linac head should consider the appropriate energy spectrum, angular and spatial distributions, and electron contamination resulting from a particular treatment head model. Over the years, some schemes that have been employed include measurement based empirical and semi-empirical models (Jiang *et al.* 2001, Fippel *et al.* 2003), phase-space-derived multi-source models (Ma et al. 1997, Ma 1998, von Wittenau *et al.* 1999, Deng *et al.* 2000, Fix *et al.* 2004), and direct use of phase-space data from MC simulations of the treatment head. Among them, phase-space sources have the potential to provide the most accurate source characterization, and they have been accepted as inputs by many of the popular CPU-based MC dose calculation tools such as DOSXYZnrc (Walters *et al.* 2005), MCNP (Seibers *et al.* 1999) and some versions of VMC++ (Gardner *et al.* 2007).

To our knowledge, phase-space source functionality does not exist in modern GPU-based MC dose engines. The vast differences in hardware architecture and simulation schemes mean that substantial and creative work is required to adopt phase-space sources in GPU-based MC dose calculation without considerable loss of efficiency. If efficiency was not of concern, the use of a phase-space file in a CPU-based dose calculation engine would be quite straightforward. Each particle could be sequentially loaded from the file, and then transported through subsequent patient-specific beam modifiers (secondary collimators and multi-leaf collimators, or MLCs), and the patient body. However, to achieve a high efficiency in a GPU-based MC dose calculation, where a large number of threads transport particles simultaneously, it is important to avoid the so-called thread divergence problem (Hissoiny *et al.* 2011, Jia *et al.* 2011). For instance, it has been observed that this problem can be greatly reduced by separating photon and electron transport and grouping particles of similar energy. Yet, since particles are usually ordered in a phase-space file randomly, their sequential use in GPU will inevitably produce thread divergence, resulting in low computational efficiency. Moreover, the amount of memory on a GPU device is limited and the overhead of transferring particle data from a large phase-space file to the GPU constitutes a significant portion of the total calculation time.

Another problem of using a patient-independent phase-space file, which stores particles at locations above all treatment plan-dependent components of a linac, is that only a small fraction of particles end up entering a patient phantom. This is because the phase-space file captures the particle fluence formed by the primary collimator and covers an area exceeding the largest possible field size, but the field sizes in a treatment plan tend to be much smaller. The overhead of reading those particles that will not eventually contribute to the dose is not significant compared to the total simulation time in traditional Monte Carlo calculations. However, it becomes a considerable issue in GPU-based computations when compared to the short time required for particle transport.

It is mainly because of the above two issues that integrating a phase-space source into GPU-based MC dose calculations is not straightforward. In this paper we present three phase-space implementations compatible with GPU-based dose calculation engines. These have been integrated into the latest version of a GPU-based MC code developed by Jia *et al.* (2011) called gDPM (now version 3.0). Previous versions of the code used a point photon or electron source model with energies sampled from a spectrum. The three methods we will present illustrate and address the challenges involved with efficient phase-space source implementations. The first method (Section 2.1) is based on the standard implementation of most CPU-based codes – transporting all particles directly from a patient-dependent phase-space file – with the minor adaptation of on-the-fly particle binning to improve the efficiency of transport on the GPU. The second method (Section 2.2) removes the requirement for patient-dependent phase-space files by introducing a simple secondary collimator model and option for using a fluence map. The third and novel method (Section 2.3) involves binning a patient-independent phase-space by particle energy, type and position prior to calculation. The sorted data structure can be then be reused in all subsequent calculations, eliminating the need for on-the-fly particle binning.

## 2. Methods and materials

### 2.1. *Method 1: patient-dependent phase-space source*

The first and simplest implementation of phase-space sources we will present has two goals: (1) enable the use of patient-dependent phase-space sources in a standard format, and (2) avoid the thread divergence issue that occurs when particles from an unsorted phase-space are transported sequentially. We support two phase-space file formats: those defined by the Electron Gamma Shower (EGS) code (Kawrakow and Rogers 2003), and the International Atomic Energy Agency (IAEA) (Capote 2007). In this method, we assume that the phase-space files were generated with the effects of the secondary collimators included. The inclusion of intensity modulation from MLCs is optional (see Section 2.4). It is necessary to launch a separate instance of gDPM v3.0 for each phase-space file (typically each field in a treatment plan is associated with one patient-dependent phase-space file). The multiple resulting dose distributions can be simply summed.

As mentioned previously, separating photon and electron simulation and grouping particles of similar energy greatly reduces thread divergence. Hence, we conduct a bin-sorting of particles from the phase-space source “on-the-fly” before launching them for dose calculations. Specifically, we allocate a number of bins that divide the particles by type and energy. Due to the similar transport mechanisms of electrons and positrons, it is not necessary to separate them, so there are only two divisions for particle type (photons, and electrons/positrons). When the CPU is sequentially loading particles from the file, each particle is placed into the energy bin corresponding to the correct particle type. Note that in gDPM, the GPU can simultaneously transport no more than N particles, where N is determined by the available GPU threads. Thus we set the maximum number of particles a bin may contain to be N. Once there are N particles in one of the bins, the bin is “full” and the particle data is moved to GPU memory and the particles are transported. This process of sorting some particles from the phase-space file and then transporting them on the GPU is repeated many times until the desired number of particles from the phase-space has been processed. At this point, some bins in the data structure may remain partially filled so these particles are also transferred to the GPU for dose calculations. In the current implementation, particle sorting and other CPU operations are performed in series (not simultaneously) with particle transport on the GPU.

If patient-dependent phase-space files are available, this method is efficient. However, the generation of patient-dependent phase-spaces using a code like BEAMnrc is time consuming and not clinically practical. In the following sections we will describe two patient-independent phase-space implementations.

## 2.2. *Method 2: patient-independent phase-space source with jaw model*

If the phase-space source is patient-independent, namely, the source is located above the secondary collimators, then gDPM v3.0 allows modeling of particle collimation and intensity modulation. Since calculation speed is of concern, we assume that transmission through the jaws is negligible and do not perform a true particle transport through the secondary collimators. Instead, simple geometrical tests are used in order to either "accept" particles that pass through the secondary collimator's aperture and transport them through the patient geometry or "reject" and terminate the particles that intersect one of the jaws. As such, the jaws are assumed to be perfect collimators with zero transmission. This approximation neglects scatter and transmission through the jaws, but these contributions to the final dose tend to be small and limited to penumbra regions. Schmidhalter et al. (2010) showed that transmission and scattering in the secondary collimators can be ignored and still achieve 95% agreement of 1%/1mm gamma criteria with EGSnrc for 6 MV and 15 MV 10x10 cm$^2$ open fields.

Intensity modulation is achieved by supplying a grid of phase-space modulation factors (fluence map) for each treatment field (Section 2.4). Phase-space particles that pass through the jaw openings are projected to the MLC plane. A weighting factor determined by the particles position in the fluence map is then assigned to the particle. If the particle already carries a weighting factor in the patient-independent phase-space, the two factors are multiplied. The final factor is carried throughout the subsequent transport process and is used to adjust the dose deposition accordingly.

As for the phase-space implementation, we take a similar approach as in the previous method. The particles passing through the jaw openings are placed in bins corresponding to particle energy and type, and particles in each bin are transported once the bin is full. We note that this approach is not optimal in the context of patient-independent phase-spaces. It requires on-the-fly sorting of the particles, which is redundant when the same phase-space file is applied to different patients or different beams of the same patient. However, the goal of Method 2 is simply to allow direct comparison of the results obtained with our simple collimation technique against those from full MC transport through the jaws so this is not of importance. In the next section we will present one method of avoiding the redundancy of on-the-fly particle sorting.

## 2.3. *Method 3: patient-independent phase-space-let source with jaw model*

We foresee that most clinical applications of GPU-based dose calculation engines will re-use a single patient-independent phase-space per linac beam model scored above the jaws. In these cases, the same phase-space data can be applied to all the beams of the same energy in a treatment plan. As in Method 2, it is sub-optimal to sort particles on-the-fly in this situation. The sorting algorithm only needs to be performed once per phase-space (not per patient) and this can be done at the model commissioning stage before any treatment dose calculations. Additionally, one might observe that a significant amount of computation time is wasted processing particles that are subsequently removed from the simulation by the secondary collimator model (described in Section 2.2). Particle transport on the GPU is very fast, hence reading data from the hard disk and transferring them to the GPU can be limiting factors on the total simulation time and should be minimized. To help resolve these issues, we have developed a phase-space sorting utility that reads the original phase-space source and divides it into separate files based on particle type, energy and position. Here we introduce the concept of *phase-space-let* (PSL). A PSL file contains a group of particles that are within the same spatial, energy and particle type bins. We avoid use of the term *beamlet* because this typically refers to phase-space divisions only by position used in the context of treatment plan optimization (Bush *et al.* 2008a). We will begin our description of this method by first discussing azimuthal particle redistribution (Bush *et al.* 2007) – this step allows reduction of phase-space latent variance and must be performed prior to PSL generation.

2.3.1. *Azimuthal particle redistribution* Patient independent phase-space files are limited by their size and usually contain too few particles to achieve small statistical uncertainty when each particle is only used once. It is therefore a common practice to recycle particles from the phase-space – a technique valid due to the subsequent random processes involved in particle transport. However, to achieve very low uncertainty in the final dose distribution, it may be necessary to recycle each particle many times. Despite the number of times a particle is recycled can be made large, the dose distribution will still be subject to phase-space latent variance (Sempau *et al.* 2001) that can not be reduced by standard particle recycling.

There are two common solutions to solve this problem: (1) generate a very large patient-independent phase-space, or (2) apply azimuthal particle redistribution (APR) or similar technique to each particle in the phase-space. The first option is often not possible, for example when the phase-space file has been downloaded from online phase-space repositories and therefore the number of independent particles in the file cannot be increased. The second option, APR, may only be applied if the phase-space is cylindrically symmetric, but this is commonly assumed to be true for modern linac design. Azimuthal particle redistribution works by assigning each particle in the phase-space new cylindrically symmetric positions and directions (Bush *et al.* 2007). That is, each particle is uniformly redistributed azimuthally (rotated about the centre of the phase-space) with its direction cosines adjusted to maintain the cylindrical symmetry of the phase-space. This allows particles to be recycled without enhancing the latent phase-space variance artefact.

The first two methods we presented utilized APR during head model simulation as a component module of BEAMnrc. For the PSL method it will generally be necessary to perform APR in order to generate high particle density PSLs. We stress the importance of APR in this section because high particle density PSLs are essential for the method to be successful. For this reason, the functionality to perform APR is included as a key component of gDPM v3.0, which is conducted at the stage of beam commissioning when generating PSLs from a phase space file.

2.3.2. *Phase-space-let generation* Phase-space-lets (PSLs) are divisions of a patient-independent phase-space based on particle type, energy and position. The divisions by type and energy serve to reduce thread divergence on the GPU. These parameters affect simulation time for each particle based on the number and complexity of interactions that take place. The last parameter characterizing a PSL, namely position, allows us to avoid spending processor time on particles that we expect to be eliminated by the secondary collimators (Section 2.3.3).

To illustrate the PSL generation algorithm, first consider a phase-space divided into two files, one containing photons only, and the other both electrons and positrons. Next, divide each of these files into $N_E$ more files, each containing particles within some energy range (for example, the first of these files will contain only particles with energy in $[0,E_{max}/N_E]$, where $E_{max}$ is the maximum energy in the phase-space file). Now divide each of these energy and type binned files into position bins, such that each position bin i contains only particles in $[x_i,x_i+dx]$ and $[y_i,y_i+dy]$, where $x_i$ and $y_i$ are the minimum x and y boundaries of the position bin that has dimensions dx and dy. For this process we adapted and expanded on the efficient phase-space sorting software published by Bush *et al.* (2008a). Note that, while the generation of PSLs is similar to the beamlet generation Bush describes, performing dose calculations using PSLs is not as straightforward and the extra dimensions (particularly energy) necessitate substantial internal changes to the dose calculation engine, as we describe in the next section. Moreover, dividing particles according to their energies in this PSL approach is conceptually different from previous beamlet approach where particles are divided only by locations. This energy dimension is of critical importance in GPU-based MC dose calculation, which allows us to transport particles with similar energy simultaneously. This is an effective method to avoid reducing computational efficiency due to a long GPU thread occurred caused by a high-energy particle.

The PSLs are stored in different files on a hard disk. For the testing cases evaluated in this work, we use particles within 20x20 $cm^2$ at the phase-space level, and divide this into 400 1x1 $cm^2$ position bins. For 400 position bins, 10 energy bins and 2 particle type bins we arrive at 8000 PSLs derived from a single patient independent phase-space. Optimal choices of the position and energy bin sizes will depend strongly on the hardware configuration in use. One may note that reading data from 8000 files is a slow process on most computer systems. However, in practice, since the field size in a treatment plan is usually smaller, only opening those PSLs near and inside the field opening (described in more detail below) reduces the processing time considerably.

2.3.3. *Dose calculations using PSLs* For our implementation utilizing PSLs in gDPM, we select PSLs that contain particles within our region of interest (ROI) defined by the field size (each field in a multi-field treatment plan may select different PSLs). Specifically, the ROI is a rectangular region determined by first finding the open field centre at the source to axis distance (SAD), and then back-projecting through the jaw openings to the plane where the PSLs are defined. A PSL is considered in subsequent dose calculation if any portion of its area is within the region of interest, as shown in Figure 1.

For each beam, the selected PSLs are assigned relative weights based on the number of particles contained. This allows us to calculate the number of particles per PSL by multiplying the total number of particles per field by each PSL weight. Consider a patient plan with $B$ fields, and take $J_i$ to be the number of PSL files *selected* for each field $i$. Then the total number of particles we will consider over all fields is

$$N_{total}^{psl} = \sum_{i}^{B} \sum_{j}^{J_i} N_j^{psl} \ ,$$

where $N_j^{psl}$ is the number of particles in the $j^{th}$ PSL file. Given the total number of particles to simulate $N_{total}^{simulated}$ and the number of photons per PSL $P_j^{psl}$, we can calculate the number of photons to simulate from each PSL $j$ in each field $i$:

$$P_{i,j}^{simulated} = N_{total}^{simulated} \frac{P_j^{psl}}{N_{total}^{psl}} \ .$$

The number of electrons to simulate per PSL with original number of electrons $E_j^{psl}$ is determined by preserving original number distribution of particles in the phase-space file:

$$E_{i,j}^{simulated} = P_{i,j}^{simulated} \frac{E_j^{psl}}{P_j^{psl}} \ .$$

Note that up to this point, calculation of the dose for planned number of monitor units (MUs) per treatment field has not been considered. To obtain a correct result, we must adjust the dose deposited by the particles assigned to each field. Recall that each particle in a phase-space file is typically given a weighting factor for dose calculations. We propagate this weighting factor $w_p$ associated with each particle $p$ into PSL files. To account for the MUs per treatment field, we calculate a new weight of each particle $w'_p$ that is simply $w_p$ multiplied by the ratio of the monitor units for the assigned field $i$, $MU_i$, relative to the total monitor units for the plan $MU_{total}$.

$$w'_p = w_p \frac{MU_i}{MU_{total}} \ .$$

Since the GPU has limited onboard memory, particle data are copied in chunks from CPU RAM to GPU global memory (as a texture). In this process, we iterate through the selected PSLs in batches corresponding to each energy division. We finish reading all the necessary particles from the PSLs in one energy division before proceeding to the next, to ensure particles of similar energy are transported together. For each PSL, particles are read sequentially and data are added to a single buffer array stored in RAM. Those familiar with GPU-based dose calculation might recognize that transporting particles grouped close in position, like in a PSL, can cause a bottleneck when many parallel threads attempt to deposit dose to the same voxel in the shared global dose matrix. This is because only a single thread can write to a particular element of a shared matrix at a time. It is therefore beneficial to transport distant particles in parallel to mitigate this memory writing conflict. To avoid introducing bias, we iterate over PSLs spatially (only reading a few particles from a PSL at a time). Additionally, the beam number is sampled randomly (for multi-field treatment plans). This spatial spread of sequential particles reduces the number of cache hits on the GPU but also reduces the chances of data writing conflicts. The first time opening each PSL file, we start reading from a random location in the file and loop to the start of the file when the end is reached. This avoids increase of latent variance when jobs are split over multiple GPUs, since it is less likely for the same particles to be re-used. After each particle is randomly and uniformly assigned to a treatment field, it is checked against the ("accept or reject") jaw model. If the particle passes the collimator acceptance criteria for the field, the particle data are added to the array. Once the array is full (corresponding to the memory limitations of the GPU hardware), the data are copied into memory on the GPU device.

Once the particle data are on the GPU, each particle receives gantry, collimator and couch rotations for the corresponding field. Its weighting factors are modified to account for the relative MUs for the assigned field and any further beam modifiers using a fluence map (Section 2.4). Finally, each particle is projected to the surface of the patient phantom and transport begins as described in Jia et al. (2011). Dose counters are updated using atomic functions. As in previous versions of gDPM, uncertainty is estimated by considering the standard deviation of dose over a number of independent batches.

### 2.4. *IMRT and VMAT simulation*

Fluence maps are required for dose calculations in intensity modulated radiation therapy (IMRT) and volumetric modulated arc therapy (VMAT) plans if the phase-space source does not already include patient-specific beam modifier simulation. Each fluence map represents cumulative transmission intensities through dynamic MLCs for one field. We use an in-house code to generate fluence maps from DICOM data. This defines a two-dimensional map for each beam angle. Each particle is projected to the fluence map plane to identify its corresponding fluence map intensity. The intensity value is multiplied by the particles weighting factor before subsequent radiation transport modeling.

### 2.5. *Dose normalization and absolute dose calibration*

The default output of gDPM v3.0 is a relative dose of Gy per initial particle, similar to DOSXYZnrc and VMC++ (Gardner *et al.* 2007). For a phase-space file generated by linac head geometry simulation, the initial particles are electrons incident on the target. To normalize the dose deposited in Gy, denoted by $D_o$, to be in units of Gy/initial particle, denoted by $D$, we need to scale by the number of initial particles that were used to generate the simulated particles. Since we generally transport a different number of particles than the total number in the phase-space, we must adjust the number of initial particles accordingly. We wish to achieve dose per electron incident on the target, so the dose is divided by the number of incident electrons use to generate the phase-space, $N_{incident}^{phsp}$. However, since we do not necessarily use each particle in the phase-space exactly once, we scale by the ratio of the total number of particles in the phase-space, $N_{total}^{phsp}$, with the number of particles actually transported (including recycling), $N_{total}^{simulated}$ .

$$D = D_o \frac{N_{total}^{phsp}}{N_{total}^{simulated} N_{incident}^{phsp}} .$$

A calibration factor is also required for conversion to absolute dose. In order to determine the calibration factor, a simulation under standard calibration conditions must be performed. The calibration factor is derived as the ratio of the calibration dose at the reference point to the MC calculated dose in Gy/initial particle at the same point.

The output of some linac models is affected by backscatter from the secondary collimators into the monitor chamber. For example, a smaller jaw setting results in a higher backscatter signal into the monitor chamber, causing the set number of monitor units (MU) to be reached sooner. Popescu *et al.* (2005) showed by Monte Carlo modelling that this backscatter decreases approximately linearly with field size, and the relative output factor (ROF) of a 40x40 $cm^2$ field compared to a 10x10 $cm^2$ field is nearly 2%. When modelling the treatment head completely, it is possible to record the backscatter into the monitor chamber and scale the final dose accordingly. However, we do not model scatter of the collimators or the monitor chamber geometry. The approach that we use in this paper is to apply backscatter factors from a table calculated through BEAMnrc simulations of various field sizes.

### 2.6. *Testing cases and hardware*

The validity of our three phase-space implementations in gDPM v3.0 is demonstrated by comparison with the BEAMnrc and DOSXYZnrc codes. Note that the underlying physics from gDPM v2.0 has not been modified. We have generated profile and depth dose curves in a homogeneous water phantom along with gamma-index tests for validation of the phase-space implementation methods. The term used in the gamma-index test to quantify spatial misalignment may not be needed when comparing doses that are perfectly registered. However, due to different modeling in geometry, e.g. jaw rejections, there could be in principle small difference between gDPM and DOSXYZnrc results. We hence prefer to use gamma-index here. Open fields for 4x4, 10x10 and 30x30 $cm^2$ fields were calculated using 150 million, 1 billion and 9 billion particles transported in the dose engine, respectively. All open field calculations were performed on a homogeneous water phantom comprised of 82x82x82 voxels with 5 mm voxel size in each dimension. The source to surface distance used was 90 cm, with a source to axis distance of 100 cm. The absolute dose calibration point was at 10 cm depth for a 10x10 $cm^2$ field, such that dose at that point is 0.774 cGy/MU. Calculation times were recorded for each of the considered methods of MC dose calculation. We also present relative output factors for various field sizes in a similar set-up.

All simulations shared the same patient-independent phase-space source from a BEAMnrc model of a 6MV Varian Clinac 21EX (Varian Medical Systems, Palo Alto, CA, USA). Recycling of approximately 20x was performed in DOSXYZnrc, and 40x in gDPM. In both gDPM and DOSXYZnrc the energy cut-offs for photons and electrons were 0.010 MeV and 0.700 MeV, respectively.

A clinical 7-field IMRT treatment was calculated using the PSL method and compared with results from the Vancouver Island Monte Carlo (VIMC) framework (Bush *et al.* 2008b). The patient phantom in this case was built with 154x90x109 voxels, each with a size of 3x3x3 $mm^3$. Approximately 500 million histories were simulated, estimated to achieve a statistical uncertainty of <1% near the isocentre.

Hardware utilized by the VIMC CPU cluster includes three compute nodes, each with 4 AMD Opteron 2.1 GHz 16 core processors, 192 GB DDR3 RAM and 7200 RPM SATA hard drives. The GPU-based system contains one AMD FX-4100 3.2GHz quad-core processor, 6 GB DDR3 RAM, Kingston HyperX SATA3 Sandforce SSD and two GeForce GTX 580 1.5 GB video cards. The GTX 580 has 16 multiprocessors with 32 cores each and a clock speed 1.54 GHz per core (NVIDIA, 2010). The CUDA version 4.0 was used.

## 3. **Results**

Depth dose and profile curves in water are shown in Figures 2 and 3, respectively, for 4x4, 10x10 and 30x30 $cm^2$ field sizes. For clarity, error bars are shown only on the results from Method 1 (BEAMnrc + gDPM). Estimated statistical uncertainty near the isocentre is less than 1% for all cases. Methods 2 and 3 are shown in comparison with the standard phase-space implementation (Method 1) and the VIMC system (BEAMnrc + DOSXYZnrc). Both Method 1 and the VIMC system used the same patient-specific phase-space generated from BEAMnrc, so we expect these results to agree well. Method 2 relies on a simplified jaw implementation that could potentially introduce systematic artefacts, but this effect was not observed in the calculations. The PSL technique (Method 3) is expected to have small systematic artefacts from (1) the simple jaw implementation and (2) the exclusion of extra-focal radiation outside selected PSLs (Section 2.3), but we suggest that the errors are tolerable for clinical treatment planning. Gamma-index tests were performed in 3D on the open field cases, and results are shown in Table 1. Over 98% of the voxels passed 2%/2mm criteria within the 10% isodose line in all cases. For the stricter criteria of 1%/1mm the success rate was over 95% for all cases. This indicates that in realistic patient plans any systematic issues from the PSL technique are likely to be within statistical uncertainty.

Calculation times of the above simulations are shown in Figure 4. The vertical time axis is put on a log scale to better observe the GPU-based results. All times are scaled to correspond to a single processor (i.e. per CPU or per GPU). The simulation time of BEAMnrc to generate the shared patient-independent phase-space is not relevant for this discussion, thus excluded. The BEAMnrc times shown are only for simulation of patient-dependent beam modifiers (jaws). Across all methods we see increased simulation time for larger field sizes - this is expected as statistical uncertainty can only be achieved across all field sizes by increasing the number of simulated particles. Since the VIMC Method (BEAMnrc + DOSXYZnrc) and Method 1 (BEAMnrc + gDPM) use the same patient dependent phase-space, the BEAMnrc portion of the calculation time is identical in both cases. The calculation times of the BEAMnrc + DOSXYZnrc for 4x4, 10x10 and 30x30 $cm^2$ field sizes were 4 + 4, 282 + 48 and 294 + 363 CPU hours, respectively (in each pair of numbers, the first is BEAMnrc, the second DOSXYZnrc). In Method 1 we see that gDPM execution time was fast, and the BEAMnrc + gDPM times were 4 hours + 16 seconds, 282 hours + 102 seconds and 294 hours + 830 seconds per CPU or GPU for the three field sizes (again in each pair of numbers, the first is BEAMnrc, the second gDPM). The high speed of the gDPM component for Method 1 is because secondary collimator simulation is performed by BEAMnrc, providing a phase-space for dose calculation that is already patient-dependent. Method 2, the jaw method, has slower gDPM execution time, but we were able to skip BEAMnrc simulation of collimating devices (and this eliminates the need for a CPU-based computing cluster). The calculation times were 73, 156 and 861 seconds for the three field sizes, respectively. Method 3, the PSL method, was the fastest of the dose calculation methods with calculation times of 17, 114 and 674 seconds for the three field sizes. The speed enhancement of Method 3 was particularly apparent for small field sizes, where a large portion of the original phase-space was excluded by the PSL selection algorithm. The PSL method also maintained a high speed for large field sizes due to pre-sorting of particles by energy and particle type (as opposed to on-the-fly sorting in Methods 1 and 2).

Relative output factors were determined for a range of field sizes to compare the PSL technique combined with simplified jaw model against BEAMnrc/DOSXYZnrc (Figure 5). The ROF percent differences between codes were less than 1% for all field sizes, within statistical uncertainty.

Finally, one clinical IMRT case was also considered. Figure 6 illustrates the agreement of the PSL method with the VIMC benchmark in the form of isodose curves. A gamma-index test showed 97% success for 2%/2mm criteria above the 10% isodose. The dose differences near the trachea result primarily from differences in electron transport in air. This region is most sensitive to particle transport physics. The PSL method completed the simulation in 50 seconds using a single CPU and GPU, while BEAMnrc + DOSXYZnrc required 2.6 + 5.8 hours on a single CPU.

## 4. **Discussion and conclusions**

We have presented three methods for phase-space source implementation in GPU-based Monte Carlo dose calculations. In particular, we have demonstrated that simulation times can be greatly decreased using our novel method for pre-processing phase-space sources, called the phase-space-let (PSL) technique. The methods were integrated into the latest version of the MC dose calculation engine gDPM v3.0. This work illustrates the trade-offs between several phase-space particle source implementations in GPU-based codes.

The simplest phase-space implementation we presented was Method 1, where a patient-dependent phase-space is generated using some other code that models linac head geometry (e.g. BEAMnrc). The patient-dependent phase-space is read by the GPU-based dose calculation engine and particles are sorted on-the-fly before transport on the GPU based on particle type and energy. This method is generally the most accurate way to use a phase-space, but linac head simulations tend to be slow and/or require access to a CPU-based computing cluster. The clear solution to this issue is to develop a GPU-based code for simulating particle transport in linac head geometry. However, this would be a sizeable undertaking both in terms of development of the code and for users to input the geometrical specifications of their head model (that are not always available). We foresee the most common usage of GPU-based calculations to be with shared patient-independent phase-space sources distributed through an online phase-space repository or commercial vendors.

With this in mind, we introduced the second phase-space implementation that instead uses patient-independent sources. The secondary collimators are approximated by rejecting particles outside the open field, and particle weights are adjusted according to a fluence map to simulate MLC modulation. The current method for modelling the secondary collimators is not a true particle transport simulation and could result in systematic error due to lack of transmission and scattering simulation. In a future investigation we may consider the inclusion of true Monte Carlo simulation of the secondary collimators, but we believe the current implementation is sufficient for typical clinical usage and allows for faster calculation times. Currently Method 2 uses the same phase-space reading code as Method 1, meaning that the source particles are sorted on-the-fly by particle type and energy. On-the-fly sorting could be avoided by adapting the code for Method 2 to use a pre-sorted phase-space if needed. The main purpose of Method 2 is to illustrate the accuracy trade-offs of an approximate jaw model. It also provides a good benchmark for comparison with Method 3.

The phase-space-let method, Method 3, involves sorting a patient-independent phase-space prior to dose calculation by particle type, energy and position. This method potentially introduces a new source of systematic error (on top of the jaw model and fluence map approximations) because PSLs outside the field dimensions are excluded from the simulation. This is different from the perfect collimator approximation because PSLs excluded from the simulation may still include some particles that would pass through the secondary collimator openings. However, if PSL selection is performed conservatively, the lost extra-focal contributions should be small with respect to the total field dose and have negligible effect on most clinical cases. Our current strategy for selecting PSLs is to project from the field centre at SAD through the field boundaries and up to the PSL plane. This should generally include most of the extra-focal contributions. Note that we could reduce the area of the selected PSLs through use of smaller PSL divisions by position. For example, using a bin size of 5x5 $mm^2$ instead of 1x1 $cm^2$ would improve the conformity of the PSL boundaries to the desired area of interest. However, whether or not this provides a performance gain will depend on the given phase-space, and more importantly on the server hardware, since the extra overhead of opening many more file pointers can be significant. Hardware considerations play a similar role in deciding the size of energy bin to use when generating the PSL data structure. Using a small bin size results in improved hardware utilization on the GPU, but the extra overhead in terms of number of file pointers and increased memory usage can be drawbacks. For our system and 6MV phase-space, we found that 1x1 $cm^2$ PSLs with 10 energy divisions proved to be a good choice. For other hardware configurations some benchmarking may be necessary to achieve the most desirable configuration. Since the treatment plan field sizes also play into this consideration, it may be optimal to generate several PSL data structures with varying parameters. Finally, a more efficient method for accessing such a large data structure could surely be implemented by the clever programmer. Another avenue for improvement includes implementation of hybrid computing, where CPU and GPU components of dose calculation are performed in parallel. Given the recent progress in GPU-based software and hardware development, we are optimistic that near-real-time MC dose calculations using phase-space sources will be achievable in the foreseeable future.

## 5. Acknowledgements

This work is supported in part by the University of California Lab Fees Research Program. The authors also acknowledge the British Columbia Cancer Foundation Innovation Support Fund and Vancouver Island Center

Medical Physics Research Fund for supporting the acquisition of hardware used in this project.

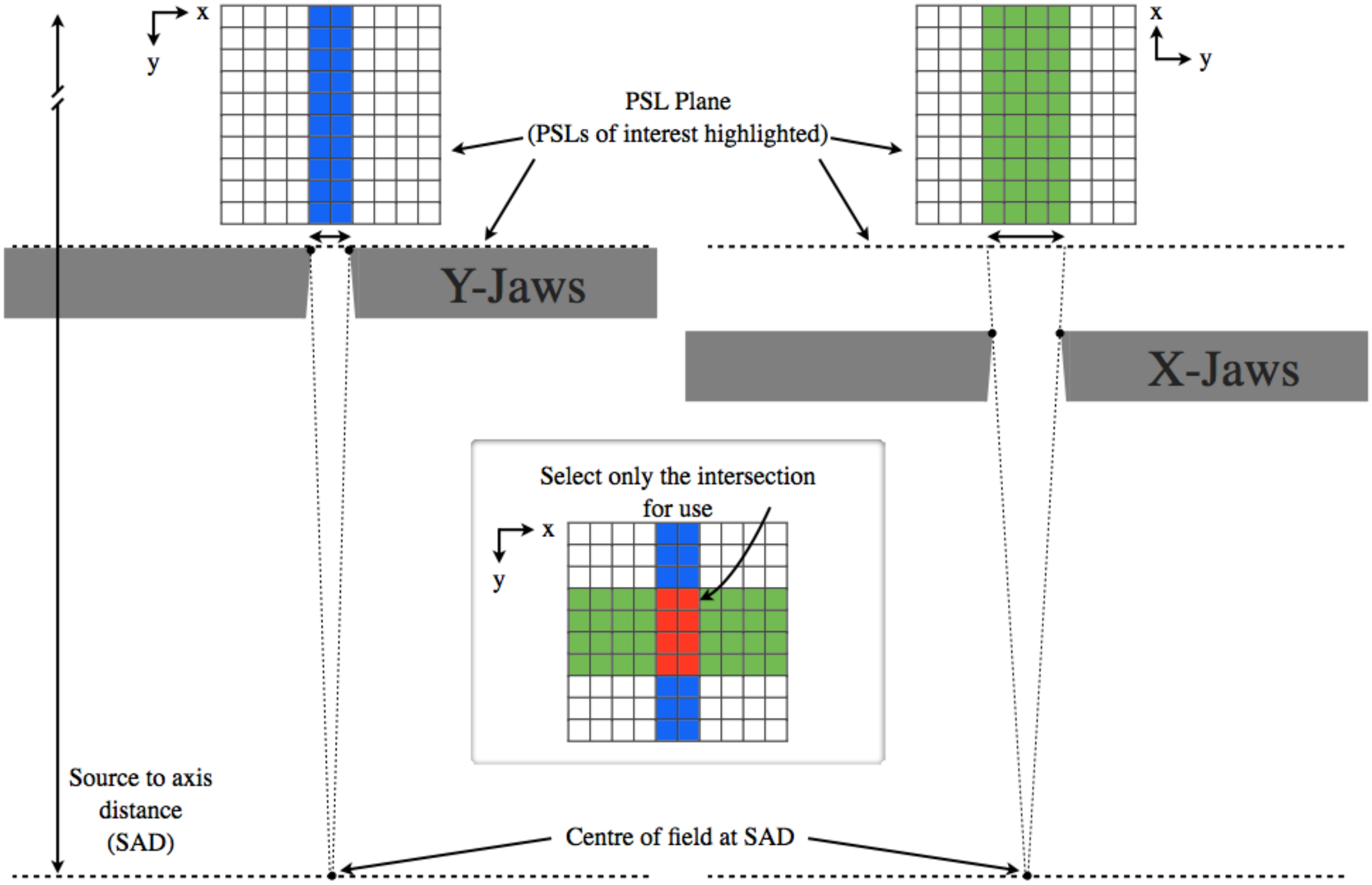

Figure 1. An illustration of how phase-space-lets (PSLs) are selected (in position space) based on the secondary collimator settings for each field (not to scale). The centre of the field at SAD is used to define the selected PSLs at the PSL plane, depending on the secondary collimator settings. The final group of selected PSLs is the intersection of those exposed by the X- and Y-jaws.

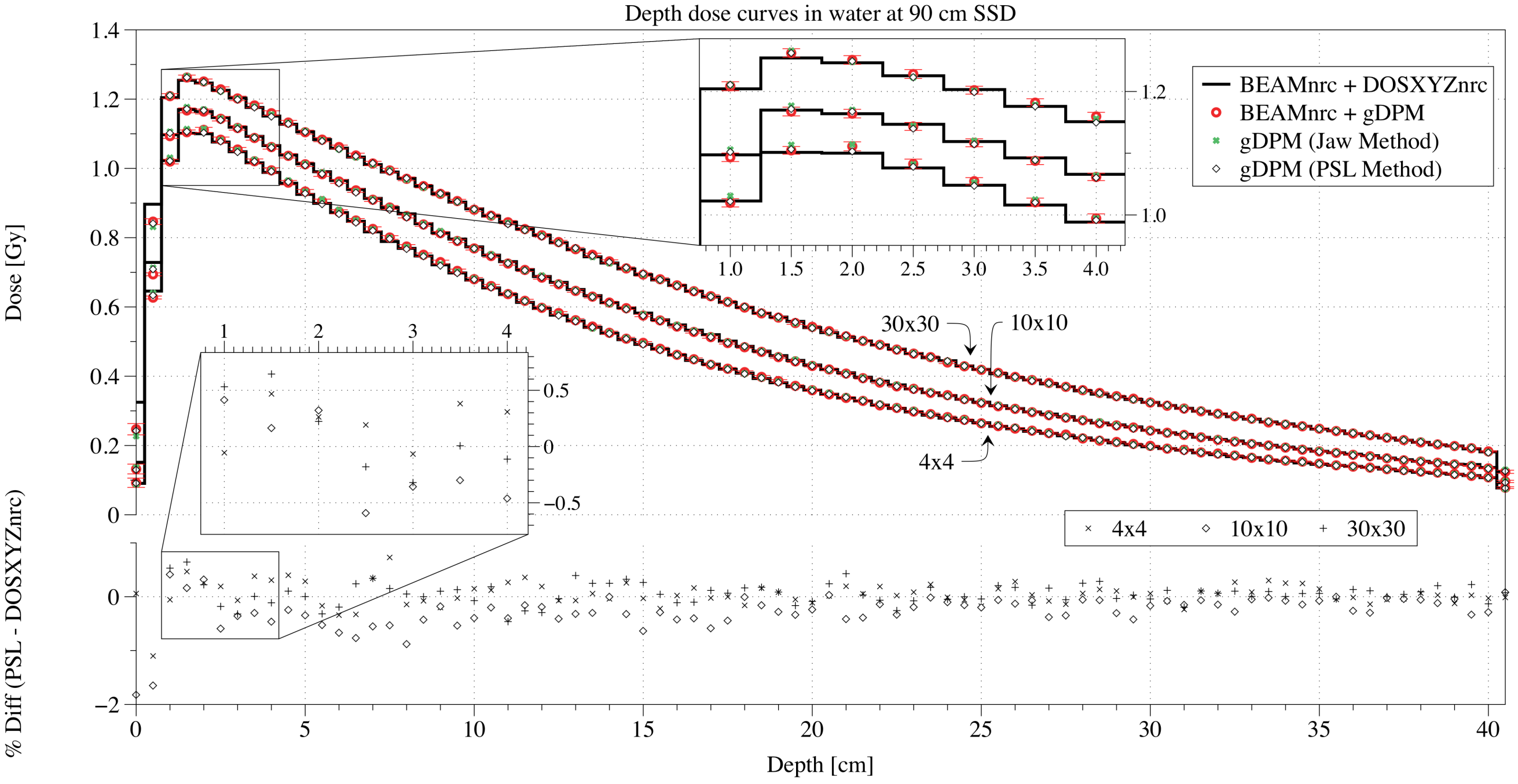

Figure 2. Depth dose curves in water for 100 MU at 90 cm SSD for 4x4, 10x10 and 30x30 cm² field sizes. The benchmark (BEAMnrc + DOSXYZnrc) is compared with Methods 1 (BEAMnrc + gDPM), 2 (gDPM Jaw Method) and 3 (gDPM PSL Method). For clarity, error bars are shown only for Method 1. The differences between Method 3 and the benchmark are shown for all field sizes, as a percentage of the maximum dose.

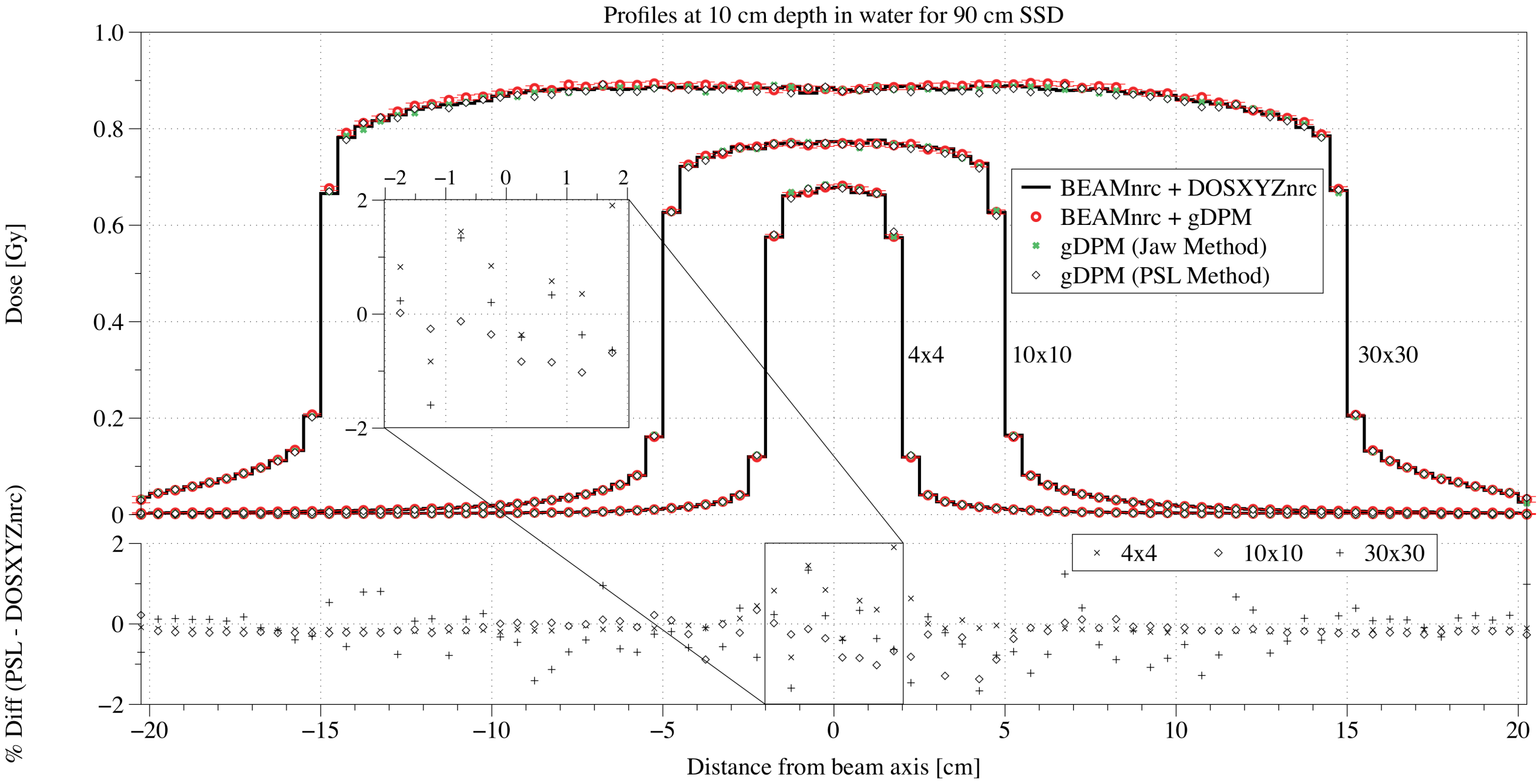

Figure 3. Cross-plane profiles in water for 100 MU at 90 cm SSD and 10 cm depth for 4x4, 10x10 and 30x30 cm² field sizes. The benchmark (BEAMnrc + DOSXYZnrc) is compared with Methods 1 (BEAMnrc + gDPM), 2 (gDPM Jaw Method) and 3 (gDPM PSL Method). For clarity, error bars are shown only for Method 1. The differences between Method 3 and the benchmark are shown for all field sizes, as a percentage of the maximum dose.

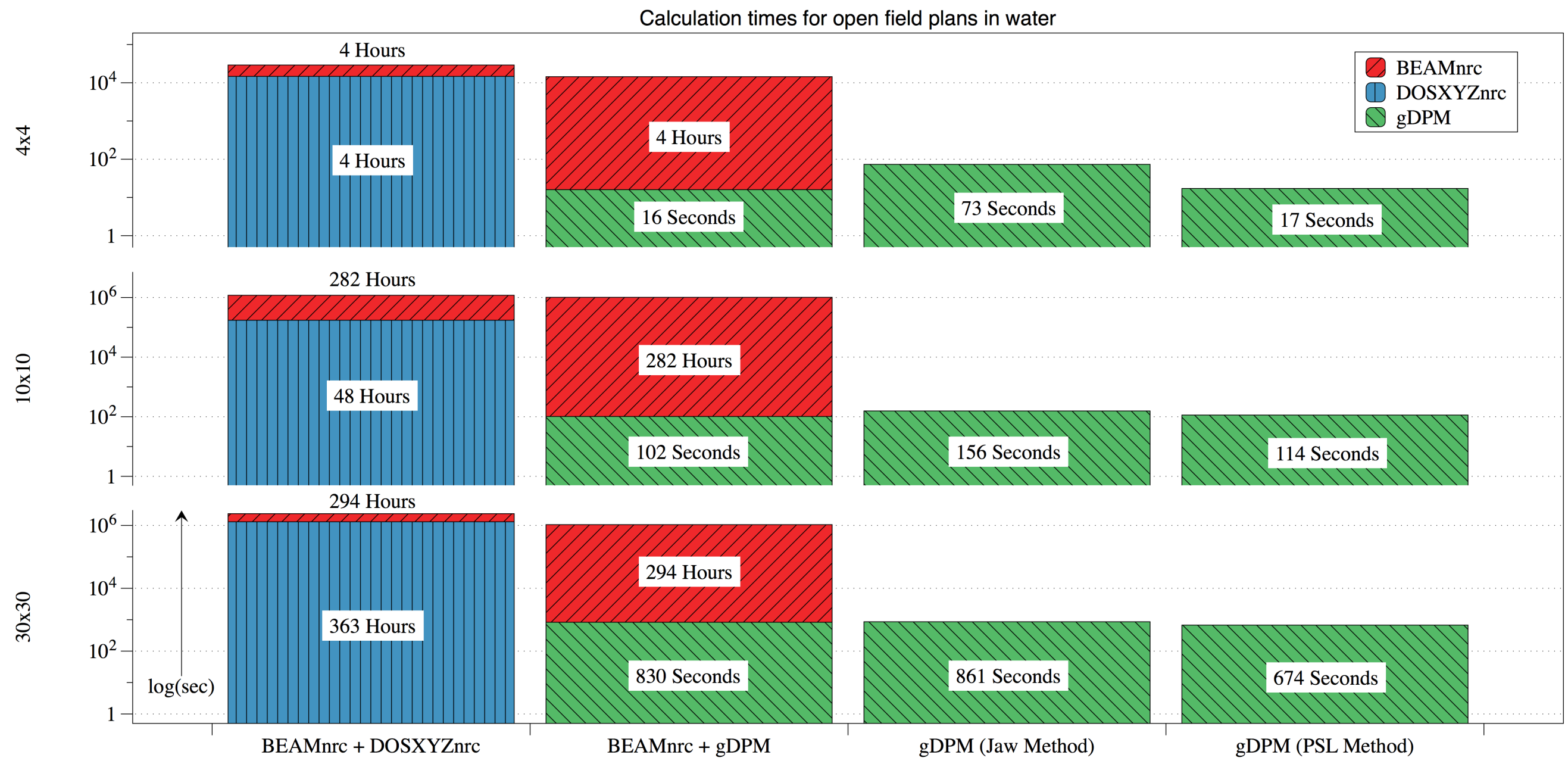

Figure 4. Calculation times for 4x4, 10x10 and 30x30 cm² open fields. The times are shown on a log scale and separated by software – BEAMnrc, DOSXYZnrc and gDPM. The benchmark (BEAMnrc + DOSXYZnrc) is compared with Methods 1 (BEAMnrc + gDPM), 2 (gDPM Jaw Method) and 3 (gDPM PSL Method). All times are scaled to correspond to a single processor (i.e. per CPU or GPU).

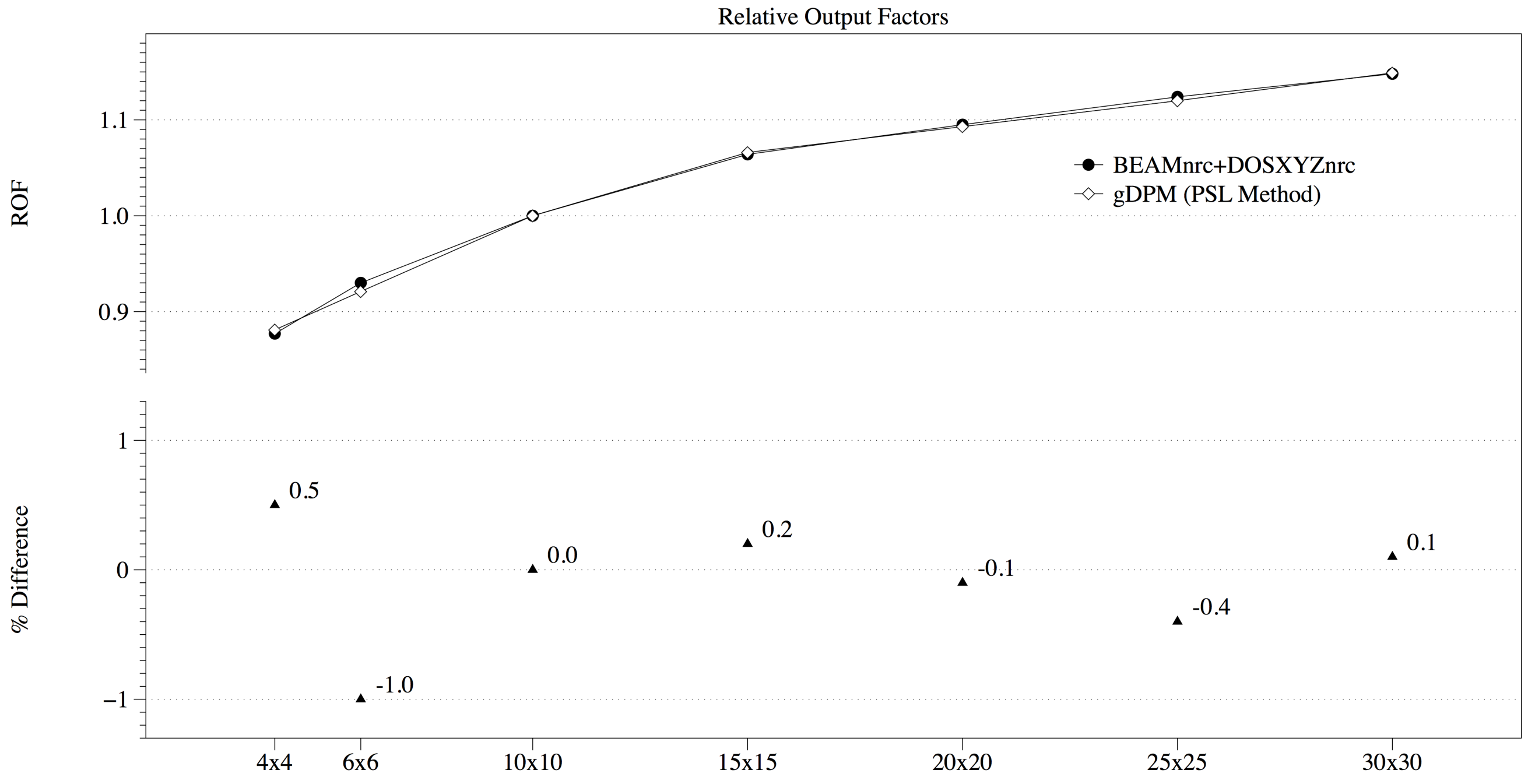

Figure 5. Relative output factors (ROFs) plotted for a variety of field sizes. The benchmark (BEAMnrc + DOSXYZnrc) is compared with Method 3, the gDPM PSL method. Percent differences between the dose engines show agreement within 1%.

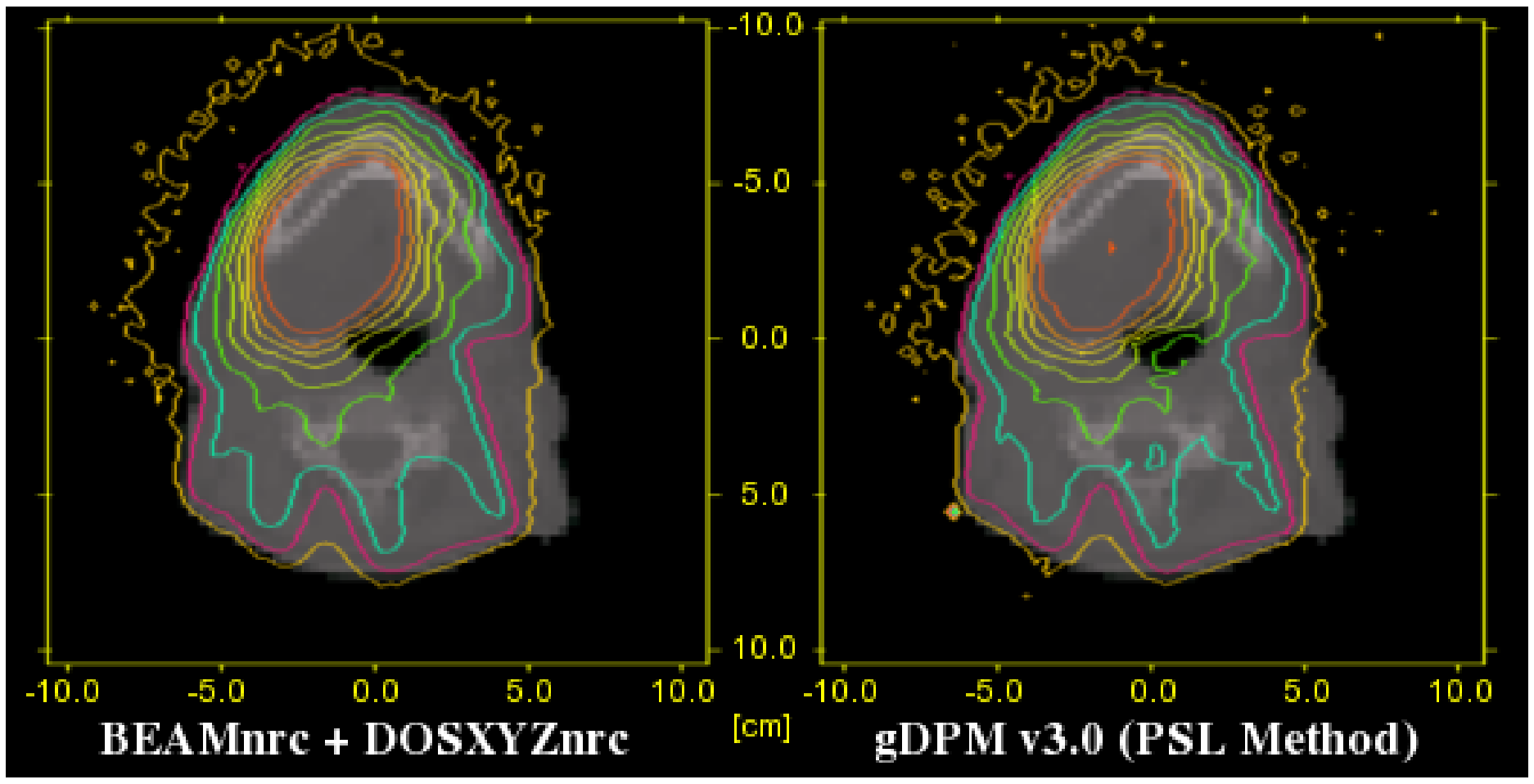

Figure 6. Isodose curves for a clinical IMRT tongue treatment, with DOSXYZnrc results pictured left and gDPM PSL results on the right. The voxel size of the phantom was 3x3x3mm³, and estimated statistical uncertainty at the isocentre was 1% for both cases.

| Field Size ($cm^2$) | Method | Gamma-index 2%/2mm | Gamma-index 1%/1mm |
|---|---|---|---|
| 4x4 | 1 | 100.00 | 99.27 |
| | 2 | 100.00 | 97.86 |
| | 3 | 99.96 | 95.41 |
| 10x10 | 1 | 99.66 | 98.99 |
| | 2 | 99.93 | 99.08 |
| | 3 | 99.92 | 98.57 |
| 30x30 | 1 | 99.61 | 98.84 |
| | 2 | 98.66 | 95.94 |
| | 3 | 98.66 | 95.26 |

Table 1. Gamma-index test results for 4x4, 10x10 and 30x30 $cm^2$ field sizes are shown. Each of the gDPM phase-space implementation methods is compared with the benchmark (BEAMnrc + DOSXYZnrc). Gamma-indices were calculated in 3D. Two criteria conditions were applied, 2%/2mm and 1%/1mm, both using data above the 10% isodose only.